\documentclass{article}

\usepackage[square,numbers]{natbib}
\usepackage[preprint]{neurips_2019}

\usepackage[utf8]{inputenc} 
\usepackage[T1]{fontenc}    
\usepackage{hyperref}       
\usepackage{url}            
\usepackage{booktabs}       
\usepackage{amsfonts}       
\usepackage{nicefrac}       
\usepackage{microtype}      
\usepackage{graphicx}
\usepackage[title]{appendix}
\usepackage{algorithm}
\usepackage{algpseudocode}
\usepackage{multirow}
\usepackage{subfig}
\usepackage{todonotes}

\title{W-Cell-Net: Multi-frame Interpolation of Cellular Microscopy Videos}

\author{%
  Rohit Saha\textsuperscript{\dag}, Abenezer Teklemariam\textsuperscript{\dag}, Ian Hsu\textsuperscript{\ddag}, Alan M. Moses\textsuperscript{\dag,\ddag,\S}\\
   \textsuperscript{\dag}Computer Science, \textsuperscript{\ddag}Cell and Systems Biology, \textsuperscript{\S}CAGEF\\
   University of Toronto \\
   \texttt{\{rohitsaha, abew\}@cs.toronto.edu},\\ \texttt{ian.hsu@mail.utoronto.ca}, \texttt{alan.moses@utoronto.ca}
}

\begin{document}

\maketitle

\begin{abstract}
    Deep Neural Networks are increasingly used in video frame interpolation tasks such as frame rate changes as well as generating fake face videos. Our project aims to apply recent advances in Deep video interpolation to increase the temporal resolution of fluorescent microscopy time-lapse movies. To our knowledge, there is no previous work that uses Convolutional Neural Networks (CNN) to generate frames between two consecutive microscopy images. We propose a fully convolutional autoencoder network that takes as input two images and generates upto seven intermediate images. Our architecture has two encoders each with a skip connection to a single decoder. We evaluate the performance of several variants of our model that differ in network architecture and loss function. Our best model out-performs state of the art video frame interpolation algorithms. We also show qualitative and quantitative comparisons with state-of-the-art video frame interpolation algorithms. We believe deep video interpolation represents a new approach to improve the time-resolution of fluorescent microscopy. Code, and models of our framework are publicly available \href{https://github.com/RohitSaha/W-Cell-Net_cellular_video_interpolation}{\textcolor{blue}{here}}.
    
\end{abstract}

\section{Introduction}
\label{section:intro}
Fluorescent microscopy is an imaging technique that is used to capture cellular and sub-cellular activities by exposing cells to light of a specific wavelength. This technique involves shining a laser at the cell, which activates the fluorophore to reveal the distribution of different molecules and proteins within the cell. However, the laser can kill or damage the cells and/or reduce  brightness (photobleaching) leading to degraded quality in future images. To circumvent this issue, researchers minimize the frequency of the cell's exposure to the laser. While this method mitigates the phototoxic effects of the laser, it limits the capture of fast cellular processes. Therefore, we propose to use deep neural networks with in-silico labeling techniques to perform interpolation, thereby generating intermediate frames that capture these fast cellular processes.

\section{Related Work}
\label{Section:rel_work}
Video frame interpolation is a long-standing computer vision problem with several applications. Classical approaches use optical flow information to generate frames at arbitrary times. However, optical flow techniques are susceptible to occlusions and motion-boundaries\cite{Barron} \cite{Herbst} . This leads to artifacts around the boundaries of moving objects in the interpolated frames. Moreover, optical flow computation is not end-to-end trainable and requires feature engineering. Recent success of deep learning in computer vision tasks such as object recognition and localization has prompted researchers to use similar Convolutional Neural Network (CNN) based 
architectures for interpolation.

    \begin{figure}[H]
        \centering
        \includegraphics[width=1.0\linewidth]{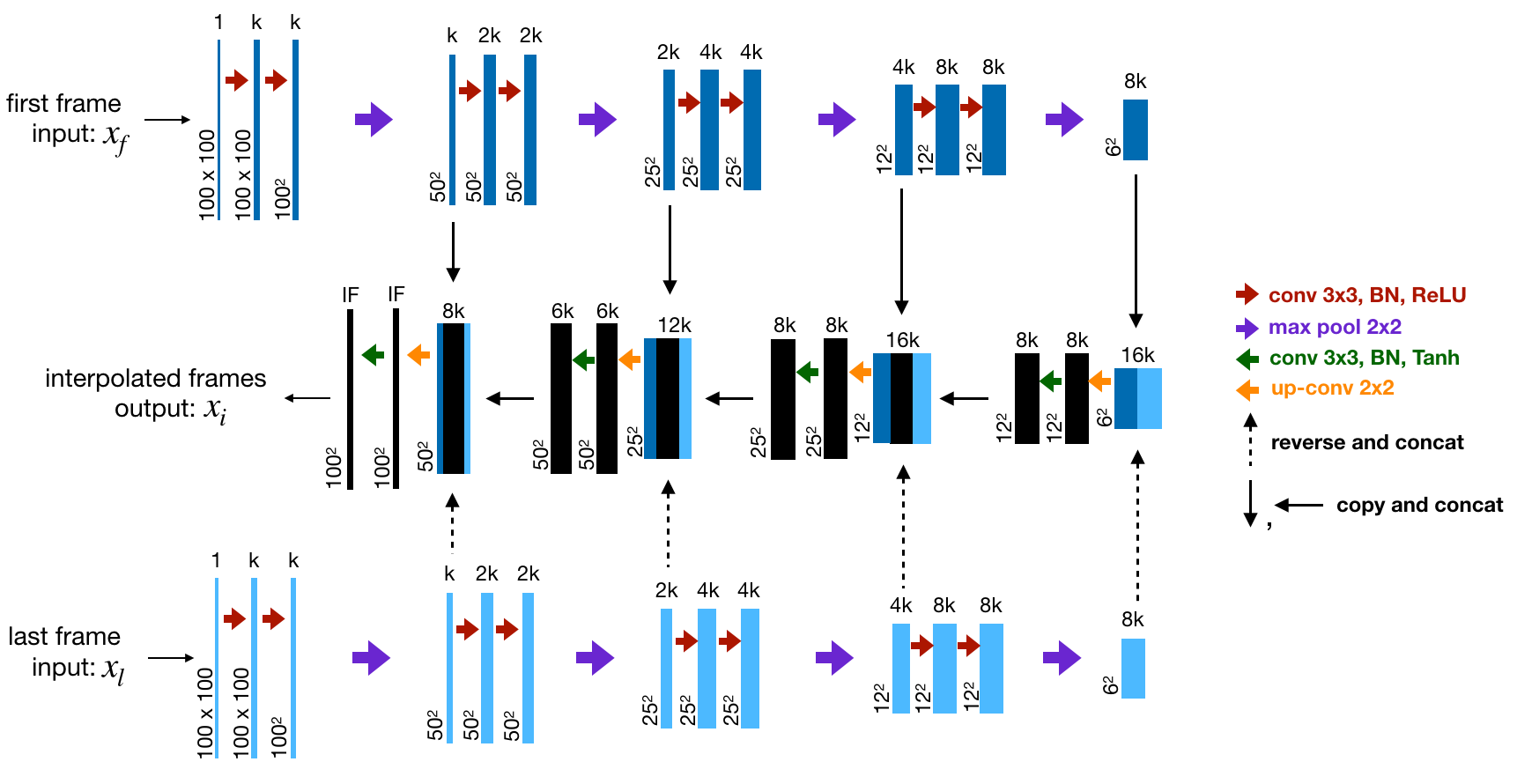}
        \caption{Proposed \textit{W-Cell-Net} architecture with Skip connections. \textit{k} controls the number of feature maps in each layer, hence the name of the architecture: \textit{W-Cell-Net-k}. \textit{IF} represents the number of intermediate frames that are being interpolated by the network. The x-y resolution is mentioned at the lower-left edge of each box. The black arrows represent copied feature maps. The dashed black arrows represent feature maps that are, first, reversed and then copied. The coloured arrows represent the different operations.}
        \label{fig:net_architecture}
    \end{figure}

 Long \textit{et al.} \cite{Long} use CNN models to learn only optical flow, which tends to make the interpolated frames blurry. Niklaus \textit{et al.}\cite{Niklaus} use local convolution over the two input frames to learn convolutional kernels for every pixel. While their technique achieves good results, it requires learning separate kernels for every pixel, which is computationally expensive. Chen \textit{et al.} \cite{Chen} propose a Bidirectional Predictive Network (BiPN) that makes use of an encoder-decoder architecture to interpolate multiple frames. In addition to using the \textit{Eucledian reconstruction loss}, they employ \textit{perceptual loss} and \textit{adversarial loss} to enforce representations that closely match the ground truth. Jiang \textit{et al.} \cite{Jiang} propose Super SloMo that employs U-Net like architecture to approximate optical flow and calculate visibility maps. First, the start and end frames are warped and linearly fused to form intermediate frames. Finally, the visibility maps are applied to the warped images, thereby excluding the contribution of occluded pixels to the interpolated intermediate frames to avoid artifacts. Moreover, computer vision and deep learning algorithms have shown tremendous promise in analyzing biological images. \cite{moen} reviews the intersection between deep leaning and cellular image analysis and surveys the field's progress in performing image classification, segmentation, augmented microscopy, and object tracking. Drawing a parallel between cellular images and videos, we believe deep learning can tackle the unprecedented problem of fluoroscent microscopy video interpolation.


\section{Methodology}
As mentinoned in Section \ref{Section:rel_work}, our model architecture is an adaptation of the network outlined by Chen \textit{et al.} \cite{Chen}. However, due to the difference in image domain between their dataset and our fluorescent microscopy images, we propose changes to their network and assess the performance of these changes. In this section, we detail the proposed model for frame interpolation. First, we detail the proposed architecture, \textit{W-Cell-Net-16} ($k=16$), and discuss the design choices. We then discuss the losses used in training our model in an end-to-end fashion. 
    
\subsection{Video Interpolation Network Architecture}
    
    The task of frame interpolation requires an understanding of object and motion appearance from both the first and last frames. Taking inspiration from Chen \textit{et al.} \cite{Chen}, we propose the \textit{W-Cell-Net} architecture, shown in Figure \ref{fig:net_architecture}, that has two encoders, $\phi_{E1}[\mathbf{x}_f]$ and $\phi_{E2}[\mathbf{x}_l]$, to process the first and last frames, $\mathbf{x}_f$ and $\mathbf{x}_l$, separately. Each encoder has four sets of convolutional blocks, each followed by a non-overlapping max pooling of size 2 x 2. A convolutional block has two sets of 3 x 3 convolution layer, each followed by Batch Normalization layer (BN) \cite{Ioffe} and ReLU activation \cite{Nair}.
    
    The decoder $\theta_{D}[\mathbf{x}_i]$, on the other hand, has four upconvolutional blocks. Each upconvolutional block is comprised of one upconvolution layer with a factor of 2, followed by a 3 x 3 convolutional layer and Tanh activation \cite{Jarrett}. At the end of every upconvolutional block, a skip connection is introduced that adds the block-wise feature maps of $\theta_{D}[\mathbf{x}_i]$ to the outputs of the corresponding convolutional blocks in $\phi_{E1}[\mathbf{x}_f]$ and $\phi_{E2}[\mathbf{x}_l]$. Skip connections help the decoder recover spatially detailed information from the two encoders, by reusing feature maps. This alleviates the issue of vanishing gradients in deep networks \cite{Zaeemzadeh}, thereby making the optimization process feasible and faster \cite{Orhan2018SkipCE}. Skip connections are implemented as concatenation operation along the channel dimension: $\left[\phi_{E1}^{(5-b)}[\mathbf{x}_f], \theta_{D}^{(b_i)}[\mathbf{x}_i], reverse(\phi_{E2}^{(5-b)}[\mathbf{x}_l])\right]$, where $b \in \{1, 2, 3, 4\}$ denotes the block number. The \textit{reverse} operation simply flips the feature maps along the channel dimension.
    
    \begin{algorithm}
	\caption{Minibatch gradient descent training of \textit{W-Cell-Net-16}. The number of blocks in the encoders and decoder is represented by \textit{B}. We used \textit{B=4}, representing four convolutional blocks in the encoders and four upconvolutional blocks in the decoder. The two encoders are denoted by $\phi_{E1}$ and $\phi_{E2}$, and the decoder is denoted by $\theta_D$.} 
    	\begin{algorithmic}[1]
    	    \State \textbf{procedure} \textsc{Train}\textsc{W-Cell-Net}
    	    \For {number of training iterations}
    	        \State Init: $\phi_{E1}LookUp, \phi_{E2}LookUp
    	        =\{\}, \{\}$
    	        \State Sample minibatch of \textit{m} samples $\{\mathbf{x}^{(1)},...,\mathbf{x}^{(m)}\}$ from data generating distribution $p_{data}(\mathbf{x})$, \newline
    	        \hspace*{2em} where $\mathbf{x}^{(j)}$ is a tuple of start frame, intermediate frames, end frame: $(\mathbf{x}^{(j)}_f,\mathbf{x}^{(j)}_i,\mathbf{x}^{(j)}_l)$.
    	        \For {\textit{b} in $\{1,2,\dots,B\}$}
    	            \State Assign: $\phi_{E1}LookUp[b]=\phi_{E1}^{(b)}[\mathbf{x}_f]$
    	            \State Assign: $\phi_{E2}LookUp[b]=\phi_{E2}^{(b)}[\mathbf{x}_l]$
    	        \EndFor
    	        \State Init: $\mathbf{\hat{x}}_i=\theta_D^{(1)}\left[concat(\phi_{E1}LookUp[4], reverse(\phi_{E2}LookUp[4]))\right]$
    	        \For {\textit{b} in $\{1,2,\dots,B-1\}$}
    	            \State Assign: $\mathbf{\hat{x}}_i=\theta_D^{(b)}\left[concat(\phi_{E1}LookUp[4-b],\mathbf{\hat{x}}_i,reverse(\phi_{E2}LookUp[4-b]))\right]$
    	        \EndFor
    	        \State Compute the loss of the model, where $\mathbf{w}=\{\phi_{E1}, \phi_{E2}, \theta_D\}$:
    	        $$ J = \frac{1}{m}\sum_{j=1}^{m}||\mathbf{x}^{(j)}_i-\mathbf{\hat{x}}^{(j)}_i||_2^2+\frac{\lambda_1}{m}\sum_{j=1}^{m}||percep[\mathbf{x}^{(j)}_i]-percep[\mathbf{\hat{x}}^{(j)}_i]||_2^2 +\frac{\lambda_2}{2}||{\mathbf{w}}||_2^2$$
    	        \State Update the parameters of $\phi_{E1}$, $\phi_{E2}$, $\theta_D$ by descending their stochastic gradient.
    	    \EndFor
        	\State The gradient-based updates can use any standard gradient-based learning rule. We used Adam Optimizer in our experiments.
        \end{algorithmic}
    \end{algorithm}
    
\subsection{Loss Functions}
    Several video frame interpolation networks \cite{FIGAN,Chen,Niklaus} utilize a weighted combinaiton of multiple loss functions to capture localized distributions while ensuring the fidelity of the global structure as well as maintaining appeal of interpolated images to human perception. Therefore, we explore the viability of loss functions of the form:
    \vspace{-0.5mm}
    \begin{equation}
        L = L_r + \lambda_1 L_p + \frac{\lambda_2}{2} L_{reg}
    \label{eq:loss}
    \end{equation}
    where $L_r$: reconstruction loss (pixel-wise loss), $L_p$: perceptual loss, $L_{reg}$: regularization loss.
    
    \textit{Pixel-wise loss}: Pixel-wise losses are useful in ensuring the global consistency of the generated images and are, therfore, used in most image synthesis and image translation machine learning models. Most notable losses are L1 loss and L2 loss. While L2 loss is the defacto cost function in many algorithms, Zhao \textit{et al.} \cite{Zhao} have shown that for image reconstruction tasks, L1 loss has better convergence properties and generates images with less ``splotchy'' artifacts.  
    
    \textit{Perceptual loss}: Supplementing pixel-wise loss with a loss calculated using high dimensional features of the generated image and the ground truth has been shown to produce images that correlate better with human perception\cite{Johnson}. Video interpolation networks \cite{FIGAN,Jiang,Niklaus} employ similar perceptually motivated losses by using the deep convolution layer outputs of a pretrained VGG network \cite{Simonyan&Zisserman}. Even though VGG is pretrained on non-medical images, Armanious \textit{et al.} \cite{Armanious} and Yang \textit{et al.} \cite{Yang} have shown that incorporating  perceptual loss using features from deeper convolution blocks of a pretrained VGG19 was useful in representing texture and style information thereby resulting in low noise and sharper images.
    
    \textit{Regularization loss}: To ensure that our model doesn't overfit, we employ weight regularization in the form of L2 weight decay as descrbed in \cite{Krogh}.

    For detailed information on the loss calculation, refer to Appendix \ref{appendix:losses}.

\section{Experiments}
In this section, we first discuss the dataset used for our experiment and provide comprehensive information about the training process as well as baselines used to assess our models. Second, we examine the performance of multiple versions of our model. Third, we compare our best model with state-of-the-art architectures designed for video interpolation tasks of naturalistic scenes.

\subsection{Dataset and Training}

    Our dataset includes 205, 0.16 fps videos of Yeast cells imaged over one hour. The Yeast cell strain was constructed from a reference strain BY4741 by integrating both the calcium reporter GCaMP3 and the yEGFP-tagged Crz1 at HO locus and at URA3 locus, respectively. The calcium reporter gene was assembled between the promoter of RPL39 and the ADH1 terminator followed with a selectable marker (LEU2). The yEGFP-tagged Crz1 was assembled between the promoter of CRZ1 and the ADH1 terminator followed with another selectable marker (URA3). Time-lapse microscopies were performed with Nikon CSU-X1 at room temperature. 488 nm laser was applied with time resolution of 6 sec/frame, exposure time of 50 msec, and 25\% laser intensity. Bright field images with out-of-focus black cell edge were acquired every minute for one experiment and every 6 seconds for another experiment. Cells were attached to glass-bottom dishes with 0.1 mg/ml Concanavalin-A as a binding agent using a standard protocol. The experimental condition was 0.2M calcium chloride solution. 
    
    Each of the 205 videos in our dataset has 600 frames. The videos can vary in cell count, that is from having only a single cell to several cells, as well as depths of imaging, higher to lower depths. Using a window size of \textit{IF}$+2$ and stride of $S=1$, where \textit{IF} $\in \{3, 4, 5, 6, 7\}$, we created five versions of our dataset. For \textit{IF}$=3$, a window of size $5$ with stride $S=1$ is slid over the video to create dataset where each entry has five frames. The first and the last frames are inputs to the model, whereas the intermediate three frames are used for training supervision. Since the dataset consists of images taken every 6 seconds, the wait time (in seconds) between the first and last frames is denoted by ``T'', where $T = 6 \times IF$. Finally, each of the five versions of the data was randomly divided into training, validation and test splits using 70-15-15 rule. Additional details of the number of samples available for each \textit{IF} is included in Appendix \ref{appendix:dataset}.
    
    All models were trained on an NVIDIA Quadro P6000 24GB GPU for a total of $100,000$ iterations. We use the Adam Optimizer \cite{kingma2014adam}, and the learning rate is initialized to be $0.001$ with default values of $beta1=0.9$ and $beta2=0.999$. For data augmentation, we randomly change the brightness and contrast of the samples. Additionally, each sample randomly undergoes both left-right and up-down flips. It is to be noted that the same degree of augmentations are applied to both first and last frames of every sample. Finally, all models are evaluated on the relevant test split version, depending on the number of frames being interpolated by the model.
    
    To our knowledge, there is no prior work relating to frame interpolation between consecutive fluorescent microscopy images. Therefore, we propose three baselines for our model. Given the first frame, last frame and desired number of intermediate frames (\textit{IF}), our first baseline, \textit{FFR}, replicates the first frame \textit{IF} times. Similarly, the second baseline, \textit{LFR}, replicates the last frame \textit{IF} times. Finally, the third baseline \textit{WF} creates \textit{IF} weighted sums of the first and last frames. \textit{FFR} and \textit{LFR} are used to inspect whether our models are naïvely replicating the first, or the last frames at the output. On the other hand, \textit{WF} approximates the characteristics of optical flow, i.e., an intermediate frame should be similar to the calculated weighted sum of its previous and next frames. Therefore, a model is preferred if its output's Eucledian distance w.r.t the ground truth is lower than the Eucledian distances calculated between the ground truth and \textit{FFR}, \textit{LFR} and, \textit{WF}. We borrow the Peak Signal to Noise Ratio (PSNR) metric from the computer vision literature of video interpolation for naturalistic scenes to evaluate the performance of all interpolation algorithms. See Appendix \ref{appendix:metrics} for details.
    
\subsection{Model Downselection}

    In this section, we perform several studies to analyze different variants of our model \textit{W-Net-Skip-k}, where $k \in \{4, 8, 16, 32, 64\}$ for the purpose of model downselection.
    
\begin{figure}[h]
        \centering
        \vspace{-10pt}
        \subfloat[]{
            \raisebox{.9\height}{
            \setlength{\tabcolsep}{2pt}
            \begin{tabular}{p{2.5cm} c c}
            \toprule
            Model & MSE $(\downarrow)$ & PSNR $(\uparrow)$\\
            \midrule 
            FFR  & 7.45 & 89.42 \\
            LFR  & 7.45 & 89.41 \\
            WF & 5.90 & 90.43 \\
            W-Cell-Net-4 & 5.64 & 90.63 \\
            W-Cell-Net-8 & 5.64 & 90.62\\
            W-Cell-Net-16 & \textbf{5.19} & \textbf{90.98}\\
            W-Cell-Net-16 (w/ $L_p,L_{reg}$) & 6.80 & 89.81 \\
            W-Cell-Net-32 & 5.30 & 90.89 \\
            W-Cell-Net-64 & 9.05 & 88.56\\
            \bottomrule
            \end{tabular}
            \label{model_performance_table}
            }
        }
        \hspace{2mm}
        \subfloat[]{
            \includegraphics[width=0.48\columnwidth]{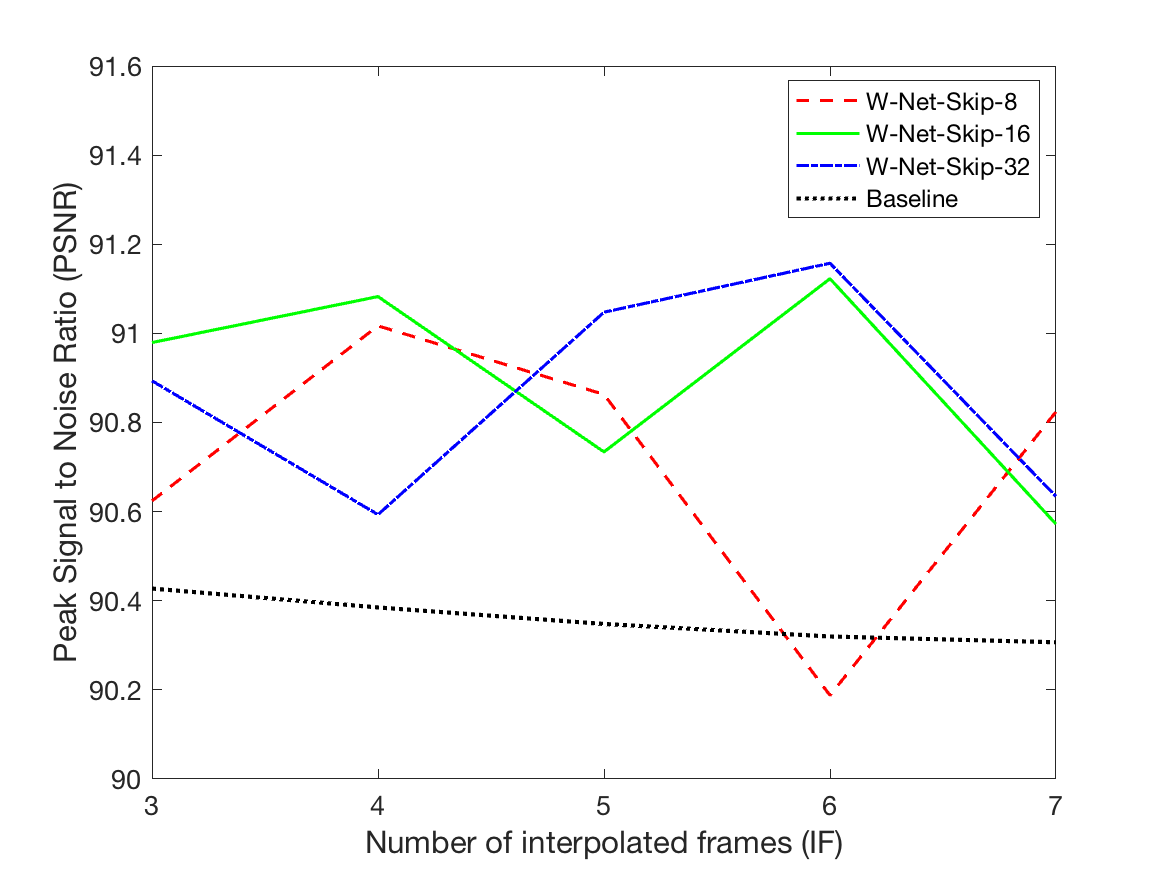}
            \label{model_performance_graph}
        }
        \caption{Performance of baselines and multiple variants of our models. (a) MSE ($\times 10^{-5}$) and PSNR values for baselines as well as our model with varying number of filter channels (\textit{k}) and added losses for the best perfoming model ( $\lambda_1=10^{-4}, \lambda_2=10^{-5}$). (b) Performance of our top three models with respect to the number of intermediate frames (\textit{IF)}.}
        \label{fig:model_performance}
    \end{figure}

    \textbf{Effect of increasing feature maps:} We first test whether a controlled increase in the number of feature maps in each layer improves the video interpolation results. Intuitively, increasing the number of feature maps, $k$, should increase the model's capacity, thereby, allowing it to learn more complex features. To this end, we trained five versions of our model with varying number of feature maps per layer. At test time, we use each model to predict three intermediate frames (\textit{IF}$=3$). Figure \ref{fig:model_performance} \subref{model_performance_table} shows that the PSNR value increases as $k$ varies from $4$ to $8$ to $16$. However, the PSNR value drops as $k$ is increased further to $32$ and $64$, making \textit{W-Cell-Net-16} the best performing model.
    
    \textbf{Effect of additional losses:} We re-trained \textit{W-Cell-Net-16} with training loss modified to include perceptual loss and weight decay regularization. Similar to the previous study, we trained the model to predict three intermediate frames. After performing a hyperparameter search, we set $\lambda_1=10^{-4}$ and $\lambda_2=10^{-5}$. Counter to our intuition, the additional losses did not improve the performance of the vanilla \textit{W-Cell-Net-16} model. We believe that the additional losses, being regularizers, over-powered the reconstruction loss, thereby reducing the model complexity and hindering the model's performance when compared to the vanilla version. Additionally, VGG features might not be useful for the task at hand \cite{alexLu}.
    
    \textbf{Impact of increasing \textit{IF}:} We also investigated the performance of some our models as we vary the number of interpolated frames. We selected the top three models from Figure \ref{fig:model_performance}\subref{model_performance_table} ($k= 8,16,32)$ and analyzed their performances w.r.t. different values of \textit{IF}. From Figure \ref{fig:model_performance}\subref{model_performance_graph}, it can be seen that the variants of the model with $k \in \{16, 32\}$ consistently perform better than the \textit{WF} baseline calculated for each \textit{IF} value. Furthermore, the model with $k=16$ has lower variance in its performance when compared with the model with $k=32$. See Appendix \ref{appendix:qual-quant} for additional plots.
    
    In summary, the vanilla \textit{W-Cell-Net-16} model consistently outperforms other variants in each of our studies.
    
\subsection{Comparison with State-of-the-art Models}
    
    We select our top two models, \textit{W-Cell-Net-16} and \textit{W-Cell-Net-32}, and compare them to current state of the art models for video interpolation \cite{Chen,Jiang}. In Table \ref{Tab:comp_stot}, we can see that both of our models outperform \cite{Chen} and \cite{Jiang}. Additionally, our best model, \textit{W-Cell-Net-16}, outperforms \cite{Jiang} with $94.94\%$ fewer parameters for both \textit{IF}$=3$ and \textit{IF}$=4$.
    
\begin{figure}[H]
        \centering
        \vspace{-5pt}
        \subfloat[First Frame] {\includegraphics[width=0.19\columnwidth]{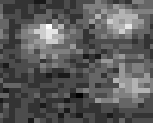}}
        \hspace{0.1mm}
        \subfloat[Last Frame] {\includegraphics[width=0.19\columnwidth]{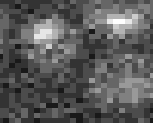}}
        \hspace{0.1mm}
        \subfloat[Ground Truth] {\includegraphics[width=0.19\columnwidth]{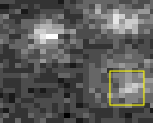}}
        \hspace{0.1mm}
        \subfloat[Super SloMo] {\includegraphics[width=0.19\columnwidth]{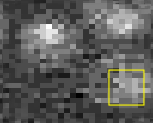}}
        \hspace{0.1mm}
        \subfloat[BiPN] {\includegraphics[width=0.19\columnwidth]{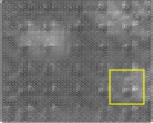}
        \label{fig:model_perf_bipn}}\\
        \subfloat[W-Cell-Net-4] {\includegraphics[width=0.19\columnwidth]{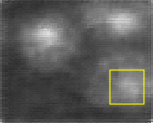}}
        \hspace{0.1mm}
        \subfloat[W-Cell-Net-8] {\includegraphics[width=0.19\columnwidth]{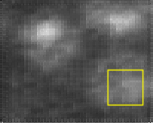}}
        \hspace{0.1mm}
        \subfloat[W-Cell-Net-16] {\includegraphics[width=0.19\columnwidth]{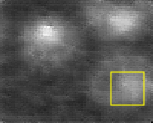}}
        \hspace{0.1mm}
        \subfloat[W-Cell-Net-32] {\includegraphics[width=0.19\columnwidth]{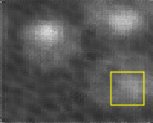}}
        \hspace{0.1mm}
        \subfloat[W-Cell-Net-64] {\includegraphics[width=0.19\columnwidth]{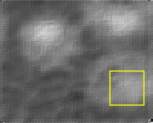}}
        \caption{Qualitative results of our model and state of the art algorithms trained to predict three intermediate frames. The predictions shown are the first of the three interpolated intermediate frames.}
        \label{fig:sample_frames}
        \vspace{-10pt}
    \end{figure}

    Figure \ref{fig:sample_frames} shows the qualitative results of our models and that of state of the art algorithms. At a first glance, and counter to quantitative results in Table \ref{Tab:comp_stot}, the sample frame generated by Super SloMo\cite{Jiang} appears to be closely matched to the ground truth relative to the other models. This could be due to the fact that the losses used in training Super SloMo focus heavily on getting an accurate and smooth optical flow approximation from the start to the end frame. Therefore, if a feature is present in the intermediate frames but missing in the start or end frames (see bounding box in Figure \ref{fig:sample_frames}), Super SloMo\cite{Jiang} might fail to capture that feature. We believe that the heavy bias in the training loss is the reason Super SloMo\cite{Jiang} fared no better than our baselines. BiPN\cite{Chen} fares poorly compared to other models as well as our baselines. This could be due to the fact that it does not have any Skip connections in its decoder, which makes optimization difficult. As a result, the interpolated frames of BiPN\cite{Chen} (Figure \ref{fig:sample_frames}\subref{fig:model_perf_bipn}) have checkered artifacts which, in-turn, lead to a lower PSNR. Additional figures with varying number of interpolated frames as well as visualization of calcium bursts \cite{Hsu} can be found in Appendix \ref{appendix:qual-quant}.
    
    It is to be noted that we did not find any existing implementation of BiPN model architecture. Therefore, we implemented the model to the best of our abilities. To make results comparable, we ensured that BiPN has the same number of feature maps in each layer as our model \textit{W-Cell-Net-16}. Additionally, we did not implement the adversarial framework of BiPN due to absence of implementation information in \cite{Chen}. 
    
    \begin{table}[h]
        \centering
        \begin{tabular}{l l c c | c c }
        \toprule
        \multirow{2}{1.7cm}{Interpolated Frames (IF)}& & \multicolumn{2}{c}{State-of-the-art} &\multicolumn{2}{c}{Ours}\\
        \cmidrule{3-6}
        & & BiPN \cite{Chen}& Super SloMo\cite{Jiang} & W-Cell-Net-16 & W-Cell-Net-32\\
        \midrule \midrule
        \multirow{2}{*}{3} & parameters & 954,138 & 24,400,819 & 1,232,698 & 4,712,026\\
        & PSNR ($\uparrow$) & 84.7814 & 89.6784 & \textbf{90.98} & 90.8935\\
        \midrule
        \multirow{2}{*}{4} & parameters & 954,332 & 24,403,992 & 1,233,148 & 4,712,604 \\
        & PSNR ($\uparrow$) & 85.6858 & 89.6927 & \textbf{91.0826} & 90.5933\\
        \midrule
        \end{tabular}
        \caption{Performance comparison of our best models to state-of-the-art video interpolation networks.}
        \label{Tab:comp_stot}
        \vspace{-25pt}
    \end{table}
    
\section{Conclusion}
We have proposed an end-to-end trainable CNN that can produce intermediate frames for fluorescent microscopy images. By reducing the exposure of cells to laser and, therefore, reducing phototoxicity and photobleaching, our model has the potential to extend the lifetime of a sample under observation. This, in turn, can reduce the amount of samples required for a study, which can translate into time-savings for cell biologists. In principle, video interpolation methods could allow ``super''-time resolution microscopy, in analogy with pixel counting methods that have facilitated spatial resolution beyond the optical defraction limit \cite{Biteen}. We show that our model, with fewer parameters, consistently outperforms the recent BiPN \cite{Chen} and Super SloMo \cite{Jiang} video interpolation architectures. We would like to briefly discuss the current limitations and use them to guide future work. First, in our studies, we perform model comparisons only for \textit{IF}$\in \{3,4\}$ due to limited compute resources. Therefore, we aim to perform a thorough analysis of different architectures on higher \textit{IF} values. Second, we realize that the noise apparent in the frames is irrelevant for most fluorescent imaging applications, and the smoothing of these images by deep models may actually be desirable. Therefore, it will be critical to evaluate the extracted quantitative information from these movies, rather than the reconstruction itself. The most important next step in development of deep interpolation methods for cellular microscopy is to demonstrate that any biases introduced by the deep learning methods are negligible relative to the biological signals.

\section*{Acknowledegment}
We would like to thank Professor Jimmy Ba at the University of Toronto for providing guidance pertaining to deep learning architectures and optimization techniques. We would also like to jointly acknowledge Dr. Tony Harris and Alex Lu for stimulating discussions on super-time resolution microscopy. We also gratefully acknowledge the support of NVIDIA Corporation with the donation of the Quadro P6000 GPU used for this research.

    
\bibliography{Report_cellular_video_interpolation}

\newpage

\begin{appendices}
\section{Dataset specifications for different \textit{IF} values}
\label{appendix:dataset}
For each version of \textit{IF}, the number of available samples were divided into train, validation and test splits using 70-15-15 rule. It is to be noted that the total samples in each row of Table \ref{tab:my_label} changes slightly due to differences in sliding window size, which is a function of \textit{IF}.

\begin{table}[!htbp]
                \centering
                \begin{tabular}{c c c c c}
                \toprule
                    \textbf{\textit{IF}} & \textbf{Total samples} & \textbf{Train samples ($70\%)$} & \textbf{Validation samples ($15\%$)} & \textbf{Test samples ($15\%$)} \\
                \midrule
                    3 & 122,179 & 85,525 & 18,326 & 18,326 \\
                \midrule
                    4 & 121,974 & 85,382 & 18,296 & 18,296 \\
                \midrule
                    5 & 121,769 & 85,239 & 18,265 & 18,265 \\
                \midrule
                    6 & 121,565 & 85,095 & 18,235 & 18,235 \\
                \midrule
                    7 & 121,360 & 84,952 & 18,204 & 18,204 \\
                \bottomrule
                \end{tabular}
                \vspace{0.2mm}
                \caption{Samples and splits available for models trained on different \textit{IF} values.}
                \label{tab:my_label}
            \end{table}

\section{Losses and Metrics Calculation}
\label{appendix:loss_metrics}

\subsection{Losses}
    \label{appendix:losses}
    The losses used for training our models include reconstruction loss ($L_r$), perceptual loss ($L_p$) and weight regularization loss ($L_{reg}$).
    \vspace{-2mm}
    \begin{equation}
        L = L_r + \lambda_1 L_p + \frac{\lambda_2}{2} L_{reg}
    \label{eq:loss_appendix}
    \end{equation}
    
    For reconsruction losses we consider pixel-wise losses, such as L1 and L2 and structural loss, such as Structural Dissimilarity loss (DSSIM).
    
    Given a batch size of ``M'', image height ``H'', image width ``W'' and number of interpolated frames ``I'', let $y_{true}$
    denote the true images and $y_{pred}$ denote the predicted images. Then the pixel-wise lossses can be calculated as:

    \begin{equation}
        Lx = \frac{1}{2}\sum_{b=1}^M\sum_{i=1}^I\sum_{h=1}^H\sum_{w=1}^W |y_{true}(b,i,h,w)-y_{pred}(b,i,h,w)|^x
    \end{equation}
    
    where $x \in \{1,2\}$ and $y(b,i,h,w) \in [0,255]$ represents the value of the pixel at row ``h'' and column ``w'' of the image that is the ``i''th frame of sample``b'' from the batch.\\ 
    
    The DSSIM loss is calculated as $DSSIM = 1 - SSIM$, where 
    
    \begin{equation}
        SSIM = \frac{1}{2MI} \sum_{b=1}^M\sum_{i=1}^I \frac{(2\mu_{y_{true}(b,i)} \mu_{y_{pred}(b,i)} + c_1) (2\sigma_{y_{true}(b,i)y_{pred}(b,i)}+ c_2)} {(\mu_{y_{true}(b,i)}^2+\mu_{y_{pred}(b,i)}^2 + c_1) (\sigma_{y_{true}(b,i)}^2+\sigma_{y_{pred}(b,i)}^2 + c_2)}
    \end{equation}
    
    \vspace{3mm}
    
    where, for our experiments, mean ($\mu$) and standard deviation ($\sigma$) are calculated using an 11x11 Gaussian filter of width 1.5. Additionally the values for the constants ($c_1,c_2$) were set to $c_1 = 4 \times 10^{-4}$ and $c_2 = 3.6 \times 10^{-3}$.\\
    
    For perceptual loss we take the L2 loss of features extracted from conv5\_3 of a pretrained VGG16 network. If we denote the conv5\_3 features of the true images by $F_{true}$ and that of the predicted images by $F_{pred}$, then the preceptual loss is,
    
    \begin{equation}
        L_{p} = \frac{1}{2}\sum_{b=1}^M\sum_{i=1}^I\sum_{h=1}^H\sum_{w=1}^W (F_{true}(b,i,h,w)-F_{pred}(b,i,h,w))^2
    \end{equation}
    
    \vspace{5mm}
    
    Weight regularization loss is implemented in the form of L2 weight decay.
    \vspace{-2mm}
    \begin{equation}
        L_{reg} = \frac{1}{2} \sum_{z=1}^Z \theta_z^2
    \end{equation}
    where $\theta$ denotes a vector consisting of all the parameters in the model and ``$Z$'' is the total number of parameters.
    
\begin{figure}[h]
        \centering
        \subfloat[L2 sample] {\includegraphics[width=\columnwidth] {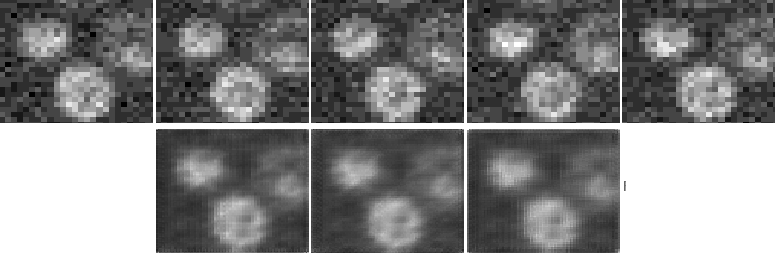}}\\
        \subfloat[L1 sample] {\includegraphics[width=\columnwidth] {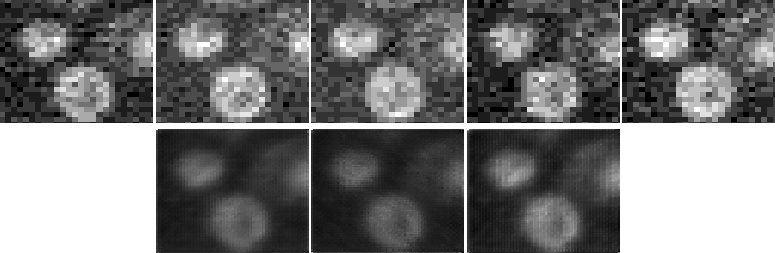}}\\
        \subfloat[SSIM sample] {\includegraphics[width=\columnwidth] {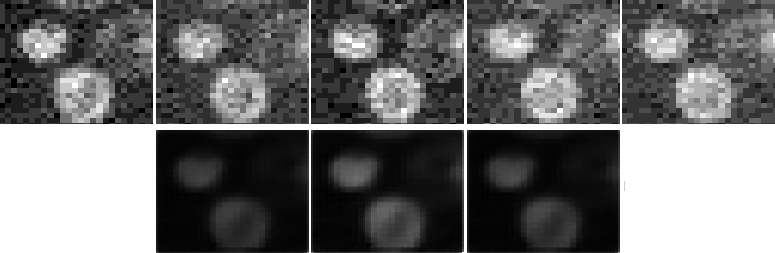}}\\
        \caption{Qualitative comparison of different reconstruction losses. Each individual image, interpolated and ground truth, is scaled in such a way as to maximize the contrast in the image. (a) Sample test frames, after $1\times10^5$ iterations , from a model trained only on L2 loss. (b) Sample validation frames, at iteration $0.95\times10^5$, from a model trained only on L1 loss. (c) Sample validation frames, at iteration $0.75\times10^5$, from a model trained only on DSSIM loss. }
        \label{fig:recon_loss}
    \end{figure}

    We chose L2 as our reconstruction loss because we observed that the predicted images tend to be brighter and higher contrast while capturing most of the cell distributions when training with L2 loss instead of L1 or DSSIM loss. While the sample frames shown in Figure \ref{fig:recon_loss} are retrived at different stages of training (due to data randomization), they capture our observations.

\subsection{Metrics}
    \label{appendix:metrics}
    The metrics used for evaluation are Mean Squared Error (MSE $\downarrow$) and Peak Signal to Noise Ratio (PSNR $\uparrow$). 
    
    For a batch size of ``M'', image height ``H'', image width ``W'' and number of interpolated frames ``I'', the metrics between the true image $y_{true}$ and predicted images $y_{pred}$ are calculated as:
    
    \begin{equation}
        MSE = \frac{1}{MHWI}\sum_{b=1}^M\sum_{i=1}^I\sum_{h=1}^H\sum_{w=1}^W (y_{true}(b,i,h,w)-y_{pred}(b,i,h,w))^2
    \end{equation}
    
    \begin{equation}
        PSNR = -10 \frac{\log{(MSE)}-\log{(255^2)}}{\log{(10)}}
    \end{equation}
    where  log x = $\log_ex$ and $y(b,i,h,w) \in [0,255]$ represents the value of the pixel at row ``h'' and column ``w'' of the image that is the ``i''th frame of sample ``b'' from the batch.
    
    For our baselines we create three functions ($f_i(S,E,t), i\in \{1,2,3\}$) which, takes as inputs the start frame (``S''), end frame (``E'') and normalized time``t'' ($t \in (0,1)$) and estimates the intermediate frame at time t. For t=0, $f_i(S,E,0) = S$ and at t=1 $f_i(S,E,1) = E$. 
    
    \subsubsection{Repeating Start or End Frames}
    This baseline would help us evaluate whether our model is learning to merely repeat the start or end frames.
    
    For repeating the start frame: $$f(S,E,t) = S, \forall t \in (0,1)$$
    
    For repeating the end frame: $$f(S,E,t) = E, \forall t \in (0,1)$$
    
    \subsubsection{Weighted Sum of Start and End frames}
    Similar to initial estimate for optical flow calculation, this baseline would help us evaluate whether our model is only learning to linearly combine the start or end frames. The weighted intermediate frame image at time t is calculated as: $$f(S,E,t) = (1-t)S + tE$$ 
    
    where, time is discretized such that $t\in\left[\frac{1}{I+1},\frac{2}{I+1} \dots \frac{I}{I+1}\right]$ and ``I'' denotes the number of intermediate frames

\section{Qualitative and Quantitative Evaluation of Multiple Models}
\label{appendix:qual-quant}

    \begin{table}[H]
        \centering
        \begin{tabular}{l c c c c c c c}
        \toprule
        \multirow{2}{*}{Model} & \multirow{2}{*}{Start channels} & \multicolumn{5}{c}{Intermediate frames } & \multirow{2}{*}{Average}\\
        \cmidrule{3-7}
        & & IF=3 & IF=4 & IF=5 & IF=6 & IF=7 & \\
        \midrule \midrule
        Best baseline & N/A & 90.4274 &	90.3850	& 90.3479 & 90.3198 & 90.3068 & 90.3574\\
        \midrule
        \multirow{5}{*}{W-Cell-Net} & 4 & 90.6271 & - &  & - & - & 90.6271\\
        & 8 & 90.6241 & 91.0169 & 90.8630 & 90.1877 & 90.8240 & 90.7031\\
        & 16 & 90.9794 & 91.0826 & 90.7340 & 91.1227 & 90.5720 & 90.8981\\
        & 32 & 90.8935 & 90.5933 & 91.0475 & 91.1574 & 90.6341 & 90.8652\\
        & 64 & 88.5636 & - & - & - & - & 88.5636\\
        \bottomrule
        \end{tabular}
        \vspace{0.2mm}
        \caption{Performance comparison of multiple models, using PSNR, with different architecture and number of filter channels on the first convolution varying from 4 upto 64.}
    \end{table}
    
\begin{figure}[h]
    \centering
    \subfloat[Interpolated frames (\textit{IF}) = 4] {\includegraphics[width=\columnwidth] {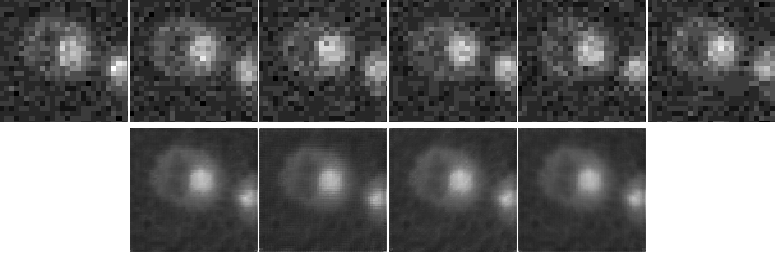}}\\
    \subfloat[Interpolated frames (\textit{IF}) = 5] {\includegraphics[width=\columnwidth] {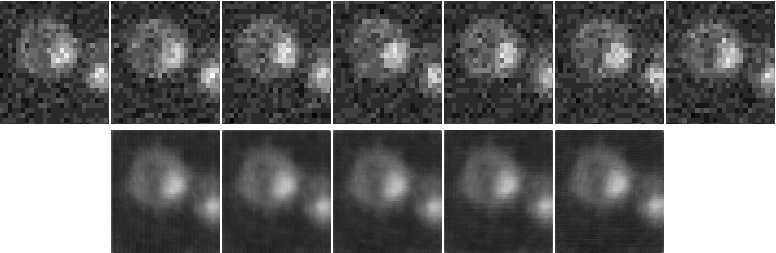}}\\
    \subfloat[Interpolated frames (\textit{IF}) = 6] {\includegraphics[width=\columnwidth] {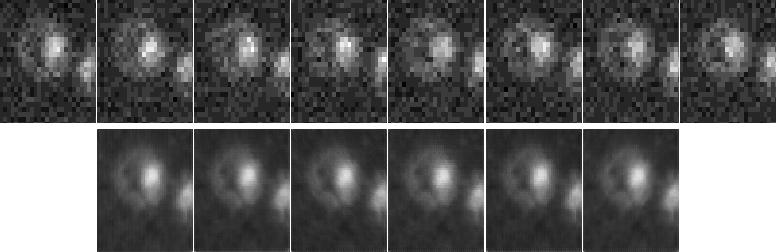}}\\
    \subfloat[Interpolated frames (\textit{IF}) = 7] {\includegraphics[width=\columnwidth] {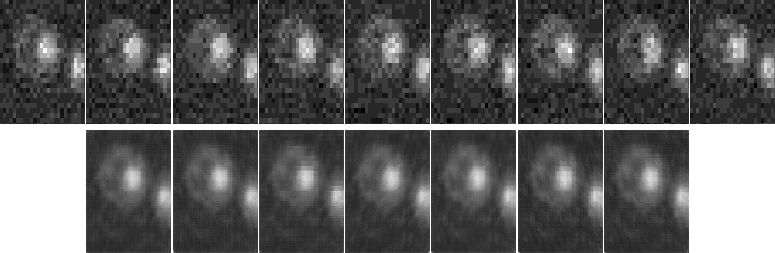}}\\
    \caption{W-Cell-Net-16 interpolation results when trained to predict different number of intermediate frames. The first row of images indicate the ground truth, whereas the second row of images are frames interpolated using W-Cell-Net-16.}
    \label{fig:W-Skip-Net-16_vs_IF}
\end{figure}

\begin{figure}[h]
    \centering
    \subfloat[] {\includegraphics[width=\columnwidth] {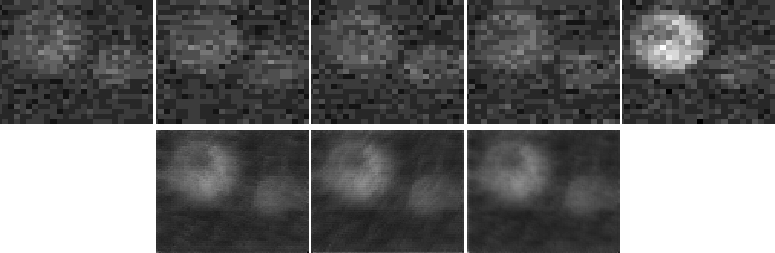}}\\
    \subfloat[] {\includegraphics[width=\columnwidth] {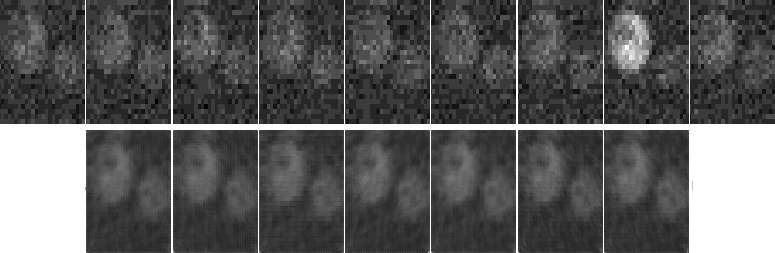}}\\
    \subfloat[] {\includegraphics[width=\columnwidth] {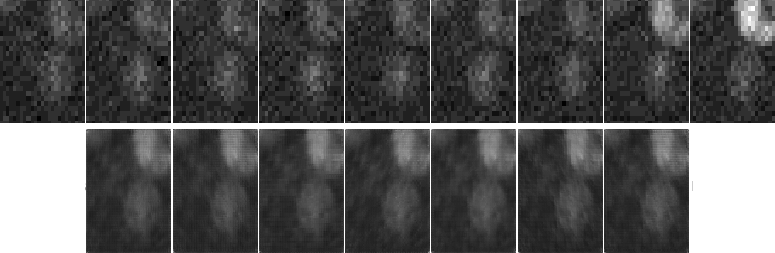}}\\
    \subfloat[] {\includegraphics[width=\columnwidth] {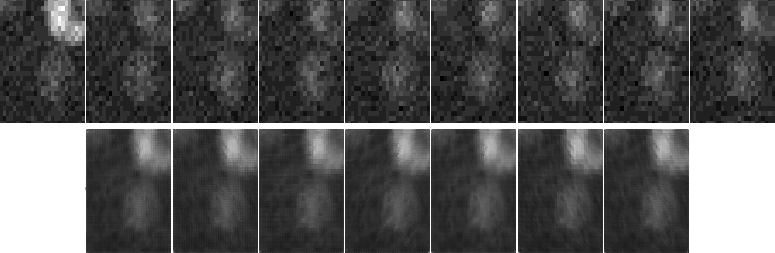}}\\
    \caption{W-Cell-Net-16 interpolation results when the difference between frames is evident to the human eye.The frames with high image contrast are a result of flurophore localizations that was prompted by calcium bursts\cite{Hsu}.}
    \label{fig:W-Skip-Net-16_bleach}
\end{figure}

\begin{figure}[h]
    \centering
    \subfloat[] {\includegraphics[width=\columnwidth] {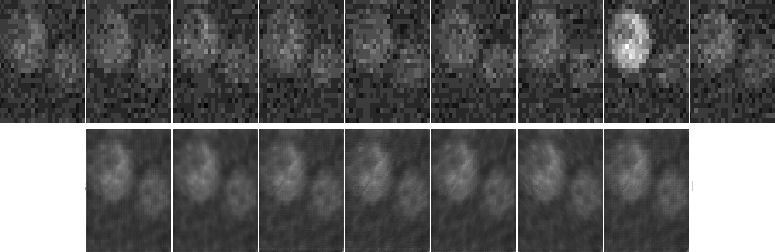}}\\
    \subfloat[] {\includegraphics[width=\columnwidth] {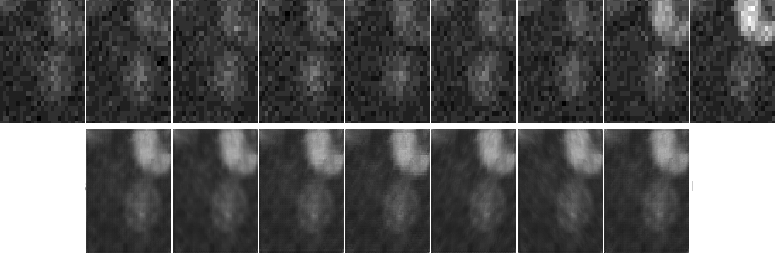}}\\
    \subfloat[] {\includegraphics[width=\columnwidth] {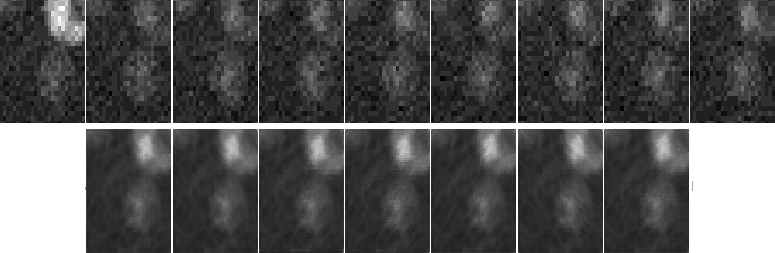}}\\
    \caption{W-Cell-Net-32 interpolation results when the difference between frames is evident to the human eye. The frames with high image contrast are a result of flurophore localizations that was prompted by calcium bursts\cite{Hsu}.}
    \label{fig:W-Skip-Net-32_bleach}
\end{figure}
\end{appendices}

\end{document}